# Electronic inhomogeneity in a Kondo lattice

E. D. Bauer[1], Yi-feng Yang[1*], C. Capan[2], R. R. Urbano[3], C. F. Miclea[1], H. Sakai[4], F. Ronning[1], M. J. Graf[1], A. V. Balatsky[1], R. Movshovich[1], A. D. Bianchi[5], A. P. Reyes[3], P. L. Kuhns[3], J. D. Thompson[1], and Z. Fisk[2]

[1]*Los Alamos National Laboratory, Los Alamos, New Mexico 87545, USA*

[2]*Department of Physics and Astronomy, University of California, Irvine, California 92697, USA*

[3]*National High Magnetic Field Laboratory, Florida State University, Tallahassee, Florida 32306, USA*

[4]*Advanced Science Research Center, Japan Atomic Energy Agency, Tokai, Ibaraki 319-1195, Japan.*

[5]*Department de Physique, Universite de Montreal, Montreal H3C 3J7, Canada*

**Inhomogeneous electronic states resulting from entangled spin, charge, and lattice degrees of freedom are hallmarks of strongly correlated electron materials; such behavior has been observed in many classes of d-electron materials, including the high-$T_c$ copper-oxide superconductors, manganites, and most recently the iron-pnictide superconductors. The complexity generated by competing phases in these materials constitutes a considerable theoretical challenge—one that still defies a complete description. Here, we report a new manifestation of electronic inhomogeneity in a strongly correlated f-electron system, using $CeCoIn_5$ as an example. A thermodynamic analysis of its superconductivity, combined with nuclear quadrupole resonance measurements, shows that nonmagnetic impurities (Y, La, Yb, Th, Hg and Sn) locally suppress unconventional superconductivity, generating an inhomogeneous electronic "Swiss cheese" due to disrupted**

**periodicity of the Kondo lattice. Our analysis may be generalized to include related systems, suggesting that electronic inhomogeneity should be considered broadly in Kondo lattice materials.**

**Classification:** Physical Sciences

Contributed by Zachary Fisk, December 19, 2010.

Electronic inhomogeneity is commonplace in materials in which strong correlations among electrons produce electronic states that compete with one another on multiple length scales(1). One early indication of such heterogeneity came from studies of the high-$T_c$ cuprate superconductors in which nonmagnetic Zn impurities were introduced into the $CuO_2$ planes of $YBa_2Cu_3O_{6+x}$ (YBCO) and $La_{2-x}Sr_xCuO_4$ (LSCO)(2); the anomalous suppression of the superfluid density of the superconducting condensate was explained within a “Swiss cheese” model comprised of normal regions around the impurity that healed over a (short) coherence length of order 20 Å within a superconducting matrix(2), later verified by scanning tunneling spectroscopy(3). Not only is superconductivity locally suppressed in the “Swiss cheese” regions, but new electronic states emerge, such as impurity resonances and other exotic forms of electronic inhomogeneity (e.g., “stripe” and “checkerboard” phases) observed in cuprates and also in other d-electron materials (e.g., manganites)(1, 4). In contrast, electronic inhomogeneity has rarely been considered in the prototypical correlated system: f-electron materials(5) in which itinerant heavy quasiparticles emerge at low temperature due to a periodic lattice of Kondo ions. In this work, we investigate the underlying electronic structure of the Kondo lattice compound $CeCoIn_5$ whose heavy quasiparticles pair to create a d-wave superconducting state below 2.3 K(6). As will be discussed, the superconductivity itself serves as a mirror that reflects the presence of electronic inhomogeneity. A thermodynamic analysis of high purity single crystals of $CeCoIn_5$, doped with different impurities ($Y^{3+}$, $La^{3+}$, $Yb^{2+}$ (7), $Th^{4+}$, Hg and Sn), reveals that lattice sites filled by these impurities create "Kondo holes"(8-9) that produce a non-superconducting component within the superconducting state, very much like the "Swiss cheese" model of the cuprates(2). Our results not only provide strong evidence for an inhomogeneous electronic ground state in this f-electron heavy fermion superconductor, they uncover fundamental properties of the Kondo lattice itself.

Substitutions for Ce (or In) in $CeCoIn_5$ by nonmagnetic elements R (or Hg, Sn) rapidly suppress $T_c$, with $T_c \rightarrow 0$ K typically in the range of 10-15% substitution for Ce (In). Figure 1 shows that, concomitant with the depression of $T_c$, there is a systematic increase in the value of C/T (T$\rightarrow$ 0 K) $\equiv \gamma_0$ that is a measure of a non-superconducting electronic contribution to specific heat in the superconducting state. In a magnetic field of H=5 T (H||c axis), the normal state Sommerfeld coefficient $\gamma_N$ follows a logarithmic temperature dependence, indicating proximity to a quantum critical point(10) for all dopants. An extrapolation of the in-field C/T data to T=0 K, such that the extrapolation conserves entropy between the normal and superconducting states at $T_c$, yields $\gamma_N$>1.2 J/mol Ce $K^2$ for all concentrations. We make the ansatz that there is an additional normal component to C/T below $T_c$ given by $\gamma_0/\gamma_N$ and compare this normal component to the reduction of the superconducting condensation energy $R_U=[U_{SC}(x)/T_c^2(x)]/[U_{SC}(0)/T_c^2(0)]$ (properly normalized relative to the condensation energy of pure $CeCoIn_5$), where $U_{SC} = \int_0^{T_c}(S_N - S_{SC})dT$. As shown in Figure 2a, the doping-induced normal state fraction comes *precisely* at the expense of the superconducting state fraction as evidenced by a common linear variation of $R_\gamma = \gamma_0/\gamma_N$ vs $1 - R_U$, for all substituents ($Y^{3+}$, $La^{3+}$, $Th^{4+}$, $Yb^{2+}$, Hg and Sn—see Fig. 2b and Fig. S1 in the Supporting Information), regardless of valence or size of the impurity atom. This unexpected result provides compelling evidence for electronic inhomogeneity in an f-electron Kondo lattice. Furthermore, the linear dependence of $\gamma_0/\gamma_N$ on impurity concentration (Fig. 2a inset) does not follow the expectation for creating electronic states in superconducting nodes through disorder in a "dirty" d-wave scenario in the strong scattering (unitary) limit (for which $\gamma_0/\gamma_N \sim x^{1/2}$) or in the weak scattering (Born) limit (Fig. 3a), implying that the impurities suppress the superconducting energy gap through the creation of intra-gap states, much like Zn impurities in YBCO and $Bi_2Sr_2CaCu_2O_{8+\delta}$ (BSCCO) (Fig. 3b)(4, 11). In this analysis, we have used the simple BCS expression for the condensation energy $U_{SC} = N(0)\Delta^2/2 \sim T_c^2$, where N(0) is the density of states at the Fermi level, to allow a comparison of the different dopants substituted into the heavy fermion superconductors. More complete calculations of $U_{SC}$ for unitary scatterers is plotted as $\gamma_0/\gamma_N$ vs $1 - U_{SC}(\Gamma)/U_{SC}(0)$ in Fig. S2 of the Supporting Information, where $\Gamma$ is the impurity scattering rate. These calculations do not reproduce the universal linear relation of $R_\gamma$ vs $1\text{-}R_U$ (Fig. 2b), furthering a scenario of electronic heterogeneity in which the dopants locally suppress superconductivity.

Our thermodynamic analysis of impurities introduced into $CeCoIn_5$ further implies that the electronic inhomogeneity arises from disruption of the coherent Kondo lattice by "Kondo holes". We estimate the characteristic energy scale of these Kondo holes through a simple binary alloy model, consistent with the creation of "Swiss cheese" holes, in which the specific heat is composed of a superconducting and normal component:

$$C_{tot} = xC_N + (1\text{-}x)C_{SC}. \qquad (1)$$

Because $C_{tot} \sim \ln(T^*/T)$ remains virtually unchanged with a Kondo lattice coherence temperature $T^* \sim 40$ K up to ~40% La in $CeCoIn_5$(12), the large contribution to electronic specific heat from these Kondo holes ($\gamma_0$~9.5 J/mol La $K^2$ for x=0.1—see Fig. 1) indicates that their effective mass is huge or equivalently that their characteristic energy scale is small, $T_{KH} = \pi R/6\gamma_0 \sim 0.3$ K for an effective 'spin-1/2' La impurity, where R is the gas constant(13); strong scattering from these massive Kondo holes leads to the loss of quantum oscillations(14), even for <1% La impurities in $CeCoIn_5$. Breaking the translational invariance of the Kondo lattice locally suppresses the superconducting gap significantly as seen in the strong reduction of the specific heat jump $\Delta C$ at $T_c$ (Fig. 3a) of doped $CeCoIn_5$, analogous to the strong temperature-dependent pair-breaking effects when Ce Kondo impurities, characterized by $T_K$ ~0.1 K are introduced into the 3.3 K s-wave superconductor $LaAl_2$(15); indeed, the suppression of $\Delta C$ in these two systems is very similar (Fig. 3a). Substitutions on the In site lead to either weaker suppression (Sn) of the gap, or stronger suppression (Cd, Hg) possibly due to additional spin-flip pair-breaking effects caused by the local nucleation of magnetism near the Cd or Hg sites(16). Further support for the local suppression of superconductivity around the Kondo holes is provided by analysis of the effective size of an impurity bound state in a d-wave superconductor(4), given by $R_{imp} = \xi_0/(1-\varepsilon^2{}_0)^{1/2}$, where $\xi_0$ = 4.9 nm is the superconducting coherence length for the $Ce_{0.9}La_{0.1}CoIn_5$ sample, determined from the initial slope of the upper critical field $dH_{c2}/dT_c$(17). The ratio of the energy of the impurity state (or resonance) to the superconducting gap $\Delta_0$ is $\varepsilon_0 = [(1-(T_{KH}/0.3\Delta_0)^2) / (1+(T_{KH}/0.3\Delta_0)^2)]$ following Ref. (18), where the strong-coupling value $\Delta_0 = 2.25T_c$ was used(6). From this formula, we find that $R_{imp}$=5.8 nm is comparable to $\xi_0$, using $T_{KH}$ =0.3 K for $Ce_{0.9}La_{0.1}CoIn_5$, consistent with local suppression of superconductivity near the La impurities. (Similar impurity length scales $R_{imp} \sim \xi_0$ are obtained for other La concentrations x=0.02 and 0.05, which have nearly identical Kondo hole energy scales $T_{KH}$ ~ 0.2 - 0.3 K(13) and values of $dH_{c2}/dT_c$ (17)). Recent scanning-tunnelling spectroscopy on Th impurities in $URu_2Si_2$(19) reveal a strong local change of the density of states in this Kondo lattice, demonstrating that Kondo holes significantly affect the normal state as well(20). Our results further strengthen the connection between the heavy fermion superconductors and the cuprates, as the suppression of $\Delta C$ of Zn-doped YBCO is similar to that of $Ce_{1-x}R_xCoIn_5$ (Fig.

3a), and also to the iron-pnictide superconductors(21-22), in which electronic inhomogeneity has been observed recently(23).

$^{115}$In nuclear quadrupole resonance (NQR) measurements further characterize the doping distribution of $Ce_{1-x}La_xCoIn_5$ and provide insight into the nature of the resulting electronic state. Figure 4 shows the NQR signal for the $4\nu_Q$ quadrupolar (±9 /2 ↔ ±7/2) transition for the in-plane In(1), as well as the temperature dependence of the spin-lattice relaxation rates ($T^{-1}_1$ ) for $x$ = 0, 0.1, and 1. The NQR peaks are relatively sharp in the pure compound (Fig. 4b) with the $LaCoIn_5$ frequency ($\nu_Q$ ~ 8.01 MHz) smaller than that of $CeCoIn_5$ ($\nu_Q$ ~ 8.17 MHz), in good agreement with previous reports(24-25). The NQR spectrum in the normal state at T = 3 K is significantly broadened for the $x$ = 0.1 sample, with the main peak (labeled A) virtually at the same frequency as in $CeCoIn_5$ and with two adjacent peaks (labeled B and C) resolved. There is no additional broadening or shift in the spectra as the sample becomes superconducting below $T_c$ = 0.9 K, confirming that the heterogeneous electronic state below $T_c$ has its origin in the normal state, as a result of doping. The lack of any intensity at the frequency corresponding to pure $LaCoIn_5$ and the similar temperature dependence of the spin lattice relaxation rates of the three peaks—all with essentially the same onset $T_c$ —rule out chemical segregation.

This thermodynamic analysis extends to other heavy fermion superconductors, as presented in Fig. 2b, and such electronic inhomogeneity may provide a framework for resolving several outstanding issues. In the 18.5 K superconductor $PuCoGa_5$, the radioactive decay of Pu-239 produces defects and/or dislocations, mimicking the “Swiss cheese” hole in $Ce_{1-x}R_xCoIn_5$. Analysis of the specific heat data of a “fresh” (~2 weeks old) and “aged” (~3 months old) $PuCoGa_5$ sample (Figs. S3 and S4 in Supporting Information), reveals that the induced normal state fraction is comparable to the superconducting state fraction as radiation damage accumulates, in agreement with self-consistent T-matrix calculations describing the rate of suppression of $T_c$(21). Furthermore, the observed anomalous reduction in superfluid density(26) with time in this strong-coupling d-wave superconductor is similar to the reduction observed in the Zn-doped YBCO and LSCO cuprates(2). Likewise, nonmagnetic Y and Th impurities(27) introduced into the exotic (odd-parity(28)) superconductor $UPt_3$ fall into this class of rather exceptional dopants. Finally, Tanatar *et al.*(29) have presented evidence for a normal component arising from an expansion of superconducting gap nodes on the Fermi surface in $Ce_{1-x}La_xCoIn_5$ ($x < 0.15$), a result interpreted within an extreme multiband model in which electrons with a small effective mass remain unpaired on small 3D Fermi surface pockets (for a different viewpoint see Ref. (30)). In contrast to this scenario, a consistent picture emerges in which impurities 1) create an inhomogeneous electronic state within the superconducting condensate of $CeCoIn_5$ (Fig. 3b) and 2) destroy the coherent Kondo lattice near the two-dimensional percolation limit ~40% for the *impurities* (not the magnetic Ce ions), corresponding to a universal scattering rate given by a resistivity of $\rho_0 \sim 35$ μΩcm(31). Novel states of matter within the "Swiss cheese" regions such as this, which do far more than just suppress the superconducting gap, are ubiquitous in cuprates(4), but our study of this particular state reveals the nature of the underlying Kondo lattice and emphasizes the delicate interplay between unconventional superconductivity and the periodic array of Kondo ions from which it originates.

Our study of $CeCoIn_5$ and other heavy fermion superconductors highlights that superconductivity itself provides a previously unappreciated window on electronic inhomogeneity in Kondo lattice materials in the form of Kondo holes. Though some theoretical ideas have been put forth to investigate the disruption of the Kondo lattice by nonmagnetic impurities(8-9), this problem remains virtually unexplored, aside from a few experimental studies(20, 32) Further investigations of Kondo holes in these f-electron Kondo lattices and superconductors, including the application of local probes such as scanning-tunnelling spectroscopy, provide an opportunity to unravel their complexity; indeed, electronic inhomogeneity in these materials may well prove to be the norm rather than the exception.

## Materials and Methods

Single crystals of $Ce_{1-x}R_xCo(In_{1-y}M_y)_5$ ($R=Y^{3+}$, $La^{3+}$, $Gd^{3+}$, $Yb^{2+}$, $Th^{4+}$; M=Sn, Cd, Hg) were grown from In flux, while single crystals of $PuCoGa_5$ were grown from Ga flux. Specific heat measurements were carried out in a Quantum Design PPMS from 0.4 K to 20 K (or from 5 K to 25 K for $PuCoGa_5$), or in a $^3He/^4He$ dilution refrigerator from 50 mK to 3 K, in magnetic fields up to 9 T. The concentrations of the impurities were determined from energy-dispersive x-ray spectroscopy (EDS). The $^{115}In$ NQR measurements were performed using a phase coherent pulsed NMR/NQR spectrometer. Several crystals with similar $T_c$ previously investigated by specific heat measurements were gently crushed into powder to improve the signal probed by the *rf* measurement. The frequency-swept $^{115}In$ NQR spectra ($I = 9/2$; $\gamma/2\pi= 9.3295$ MHz/T) were obtained using an auto-tuning probe in a $^3He$ cryostat. The spectra were obtained by stepwise summing the Fourier transform of the spin-echo signal. The values of the spin lattice relaxation time $T_1$ were obtained by fits of the recovery of the nuclear magnetization M(t) after a saturation pulse. The self-consistent T-matrix calculations of the specific heat jump are described in detail in Ref. (33).

In our thermodynamic analysis, estimates of the normal state electronic specific heat coefficient $\gamma_N$ were obtained by linear extrapolation to T=0 K (a determination of $\gamma_N$ obtained by fits of the data to the model of Moriya and Takimoto(34) for critical fermions yields similar values within 5-10%) and requiring entropy balance at $T_c$. The superconducting fraction of the heavy electrons is calculated by $R_U^*=U_{sc}/T_c^2/E_c$, where $T_c$ is determined by equal entropy construction above and below $T_c$, $U_{sc}$ is the superconducting condensation energy determined from the integration of the entropy difference between the normal (high field) and superconducting (zero field) states up to $T_c$, and $E_c$ is a constant described below. If the high field data were not available for all doping levels, we used the normal state value for $\gamma_N$ of the pure compound as an approximation for the doped compounds; the insensitivity of C/T just above $T_c$ for all concentrations (Fig. 1) and the entropy balance at $T_c$ indicate this approximation is reasonable. In a few cases (x=0.05 La, 0.012 Sn, 0.03 Yb), the entropy balance was not satisfied in all available data sets (see Fig. 1); therefore, we added a small linear term (~3-5% of $\gamma_N$) to correct the entropy balance and obtain two different values of $U_{sc}$ before and after the correction for a better relative comparison within each doping series. We take the average and their difference gives the error bars for $R_U$ of these samples in Figure 2.

In our analysis of $Ce_{1-x}R_xCo(In_{1-y}M_y)_5$, we have taken approximately $E_c=U_{sc}(0)/T_c(0)^2$ from the pure compound $CeCoIn_5$. This, together with the experimental fact $R_\gamma+R_U=1$ in Figure 2, indicates that the pure compound of $CeCoIn_5$ has negligible amount of impurities (i.e., comparing $\gamma_0$=0.04 J/mol-Ce K$^2$ to $\gamma_N$ = 2 J/mol-Ce K$^2$). However, in some heavy fermion compounds such as $U_{1-x}M_xPt_3$ and $PuCoGa_5$, even the pure compounds have a significant number of defects or a large intrinsic $\gamma_0$. In this case, it is necessary to define a parameter $E_c$, which corrects the normalization by $U_{sc}(0)/T_c^2(0)$ for additional disorder and/or systematic errors (see Table 1 of the Supporting Information), to best fit all the doped data of a given system. For example in $PuCoGa_5$, $R_\gamma$~0.2 even in the "fresh" sample, suggesting a large amount of defects caused by radiation damage, which is consistent with theoretical calculations that indicate $T_c$=19.1 K in an undamaged material and account for the decrease in $T_c$ with time(21, 26). In the case of $UPt_3$, the pure material has a different condensation energy from the doped compounds (Fig. 2b), reflecting the double superconducting transition and also the sensitivity of this exotic odd-parity superconductor to impurities.

* Present address: Beijing National Laboratory for Condensed Matter Physics and Institute of Physics, Chinese Academy of Sciences, Beijing 100190, China

**Acknowledgements:** Z.F. would like to thank Ilya Vekhter, Piers Coleman, and Lev Gor'kov for useful discussions and the hospitality of the Aspen Center for Physics. E.B. and J.T. would like to thank H. Yasuoka for helpful discussions. This work was performed at Los Alamos National Laboratory under the auspices of the U.S. Department of Energy, Office of Basic Energy Sciences, Division of Materials Sciences and Engineering. Work at UC-Irvine was performed under the National Science Foundation grant NSF-DMR-0801253.

**Author contributions:** Both E.B. and Y.Y. contributed equally to this work and wrote the manuscript. A.D.B., C.C., and E.B. prepared samples. A.D.B., A.R., C.C., C.M., E.B., F.R., H.S., P.K., R.M., R.U., and Y.Y. collected and analyzed data. A.V.B. and M.G. performed theoretical calculations. Z.F. and J.T. designed the experiments and contributed to the interpretation of the data.

**Competing interest statement:** The authors declare that they have no competing financial interests.

**Corresponding author:** Eric Bauer, Los Alamos National Laboratory, MS K764, Los Alamos, New Mexico 87545. Phone: 505-665-0942. Fax: 505-665-7652. Email: edbauer@lanl.gov

## Figure Legends

**Figure 1. Specific heat coefficient and entropy of $CeCoIn_5$ when nonmagnetic impurities (La, $Yb^{2+}$) replace Ce or when In is replaced by Sn or Hg in the crystal lattice.** Specific heat, plotted as C(T)/T (lower panel), and entropy S(T) = ∫(C/T)dT (upper panel), of $Ce_{1-x}R_xCoIn_5$ **a)** R = $La^{3+}$ (Ref. (29)) and **b)** R = $Yb^{2+}$ and **c)** $CeCo(In_{1-x}Sn_x)_5$ showing the suppression of superconductivity and the increase of the residual superconducting state specific heat coefficient $\gamma_0$ determined from a linear extrapolation of the C/T data to T=0 K, consistent with a superconducting gap with lines of nodes observed previously(6). The normal state Sommerfeld coefficient $\gamma_N$ was determined from a linear extrapolation of the C/T data to T=0 K that balances entropy between the normal and superconducting states as shown in the upper panels of **a)**, **b)**, and **c)**. The dashed lines in the lower panel of **a)** are an example of the extrapolation of the C/T data used to determine $\gamma_0$ and $\gamma_N$ for the $Ce_{0.95}La_{0.05}CoIn_5$ sample.

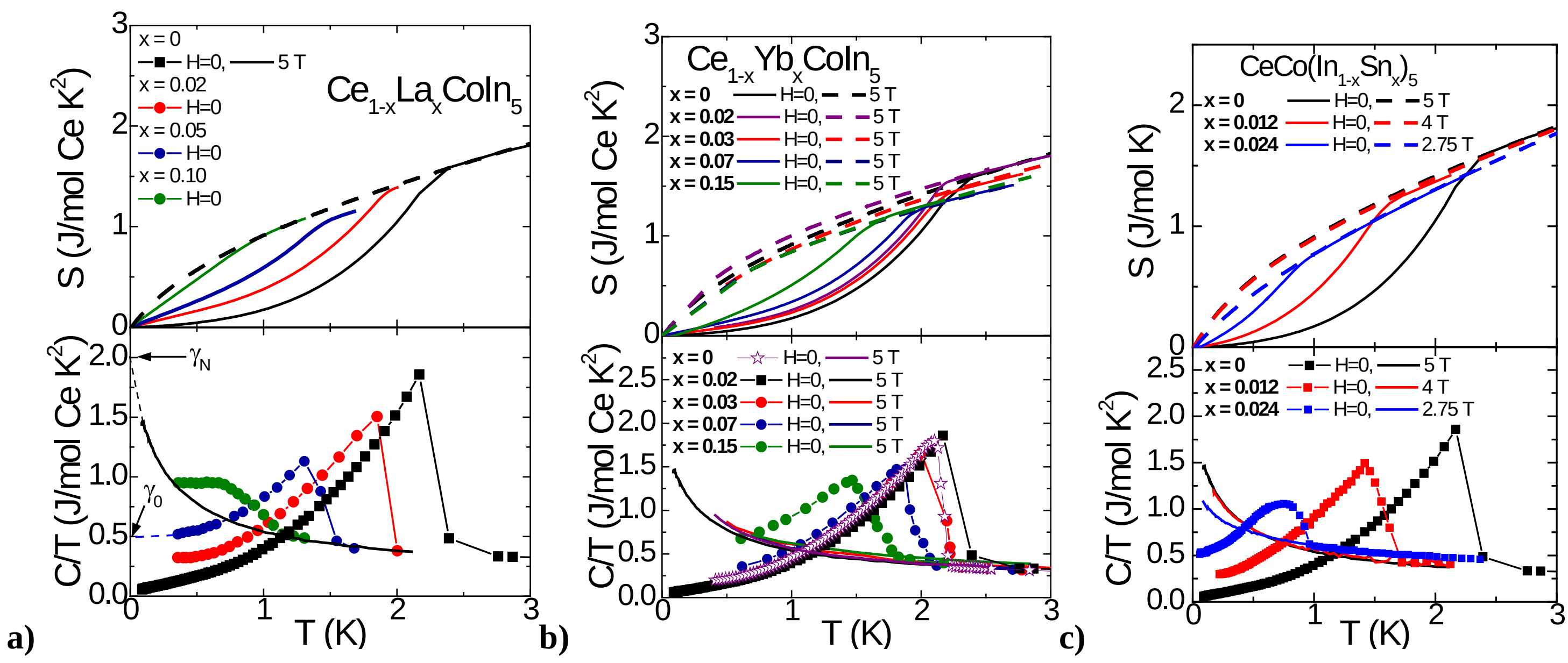

**Figure 2. Normal state fraction of the inhomogeneous heavy fermion ground state compared to the superconducting fraction of $CeCoIn_5$ with the introduction of nonmagnetic impurities and other heavy fermion systems at zero temperature.** **a)** Normal state fraction of the inhomogeneous heavy fermion ground state $R_\gamma = \gamma_0/\gamma_N$ (determined from the C/T data in Figure 1) of $Ce_{1-x}R_xCo(In_{1-x}M_x)_5$ (R = $La^{3+}$, $Yb^{2+}$; M=Sn, Hg) and superconducting state fraction, $1 - R_U$, where $R_U = [U_{SC}(x)/T_c^2(x)]/[U_{SC}(0)/T_c^2(0)]$, obtained from the superconducting condensation energy $U_{SC} = \int_0^{Tc}(S_N - S_{SC})dT$ (properly normalized relative to the condensation energy of pure $CeCoIn_5$). The linear relation between the two fractions indicates the superconductivity is excluded from a volume surrounding the impurity atom. Inset: Linear variation of $\gamma_0/\gamma_N$ as a function of impurity ($La^{3+}$, $Yb^{2+}$, Sn, Hg) concentration *x*. Error bars for $R_\gamma$ were obtained from a linear least-squares fits to the superconducting and normal state C/T data, while error bars for $R_U$ were obtained from uncertainties in values of $U_{SC}$ subject to entropy balance at $T_c$. In some cases, systematic errors in the entropy were corrected for using a procedure described in the Methods Section. **b)** Same as in **a)** including the unconventional superconductors $U_{1-x}M_xPt_3$ (M = Th, Y) and $PuCoGa_5$ (radiation damage induced impurities), $CeCoIn_5$ and $Ce_{0.994}Gd_{0.016}CoIn_5$ in magnetic field, showing the general applicability of the analysis. The superconducting state fraction, $1 - R^*_U$, where $R^*_U = R_U/E_c$, with $E_c$ comprising a small normalization factor to account for the $\gamma_0$ of the pure material as explained in the Methods Section, is linear with respect to the normal state fraction indicating the superconductivity is excluded from a volume surrounding the impurity atom, which implies an inhomogeneous electronic state. The parameters used in the analysis are given in Table 1 of the Supporting Information. The lines in **a)** and **b)** are guides to the eye.

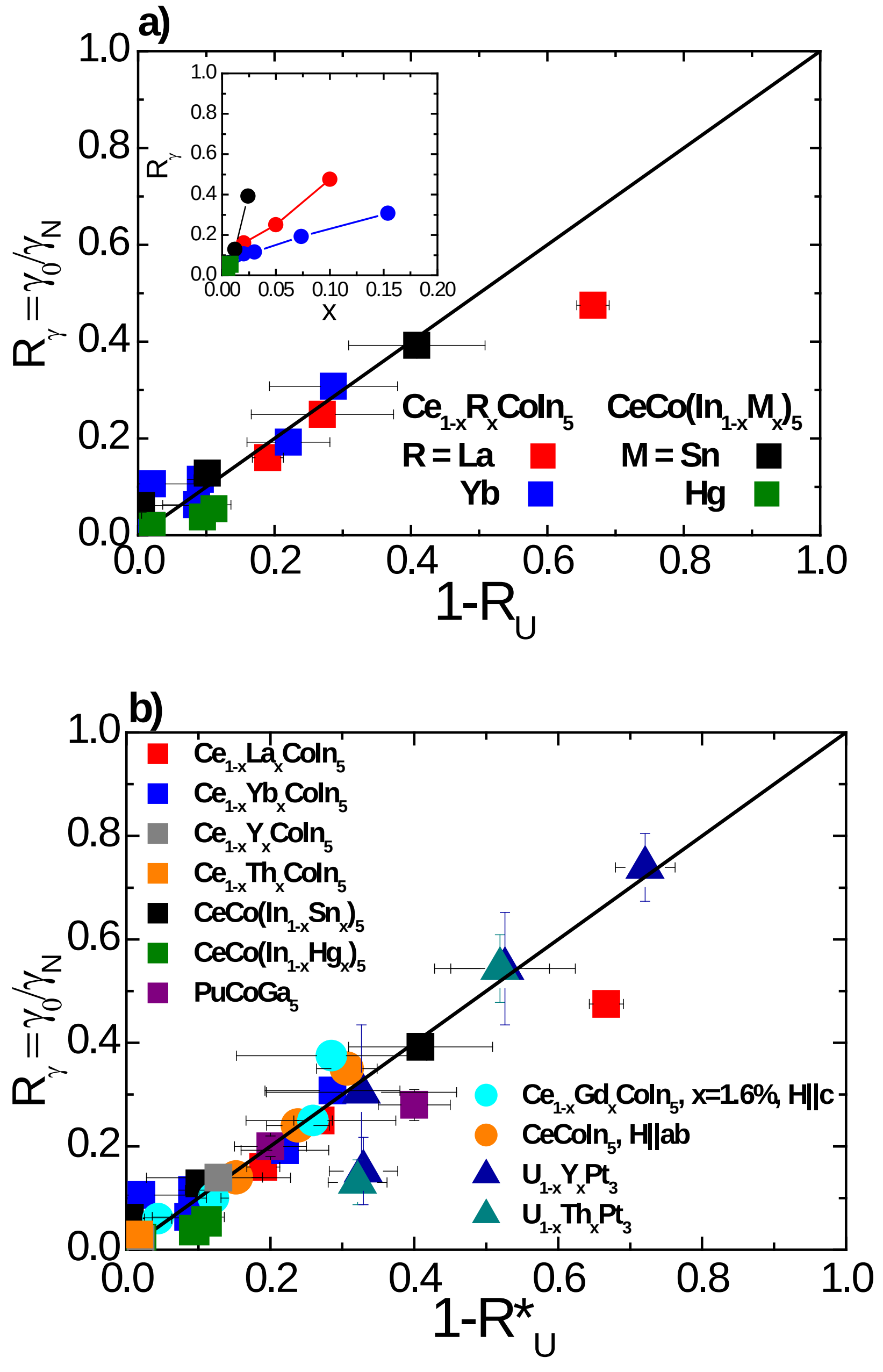

**Figure 3. Superconducting specific heat jump $\Delta C$ vs $T_c$ of $Ce_{1-x}R_x(In_{1-y}M_y)_5$ and $La_{1-x}Ce_xAl_2$, normalized to the values $\Delta C_0$ and $T_{c0}$ of the pure materials $CeCoIn_5$ and $LaAl_2$, and schematic of various novel inhomogeneous electronic states produced by impurities in cuprates and in $CeCoIn_5$. a)** Normalized superconducting specific heat jump $\Delta C/\Delta C_0$ vs normalized transition temperature $T_c/T_{c0}$ of $Ce_{1-x}R_x(In_{1-y}M_y)_5$, $La_{1-x}Ce_xAl_2$ (Ref. (15)), and Zn-doped YBCO (Ref. (11)). The black line is the Abrikosov-Gor'kov (AG) calculation for an s-wave BCS superconductor with magnetic impurities, while the blue and red lines are self-consistent T-matrix calculations for a d-wave superconductor with strong (unitary) or weak (Born) nonmagnetic scattering, respectively. The dashed lines are guides to the eye. Error bars for $T_c$ for $Ce_{1-x}R_x(In_{1-y}M_y)_5$ were determined from the 10% and 90% values of $\Delta C/T$, and uncertainties in $\Delta C$ were determined from uncertainties in the entropy-conserving equal area construction. **b)** Schematic of local density of states vs energy E near the impurity showing three possible exotic forms of electronic inhomogeneity that emerge as intra-gap states of the unconventional superconductor, including 1) an impurity resonance located near E = 0 due to a strong scatterer (e.g., Zn impurities in BSCCO), 2) a resonance away from E = 0 due to an intermediate or weak scatterer (e.g., Ni impurities in BSCCO) and 3) proposed heavy Kondo hole (red arrow) that disrupts the superconducting $CeCoIn_5$ Kondo lattice (black arrows).

. .

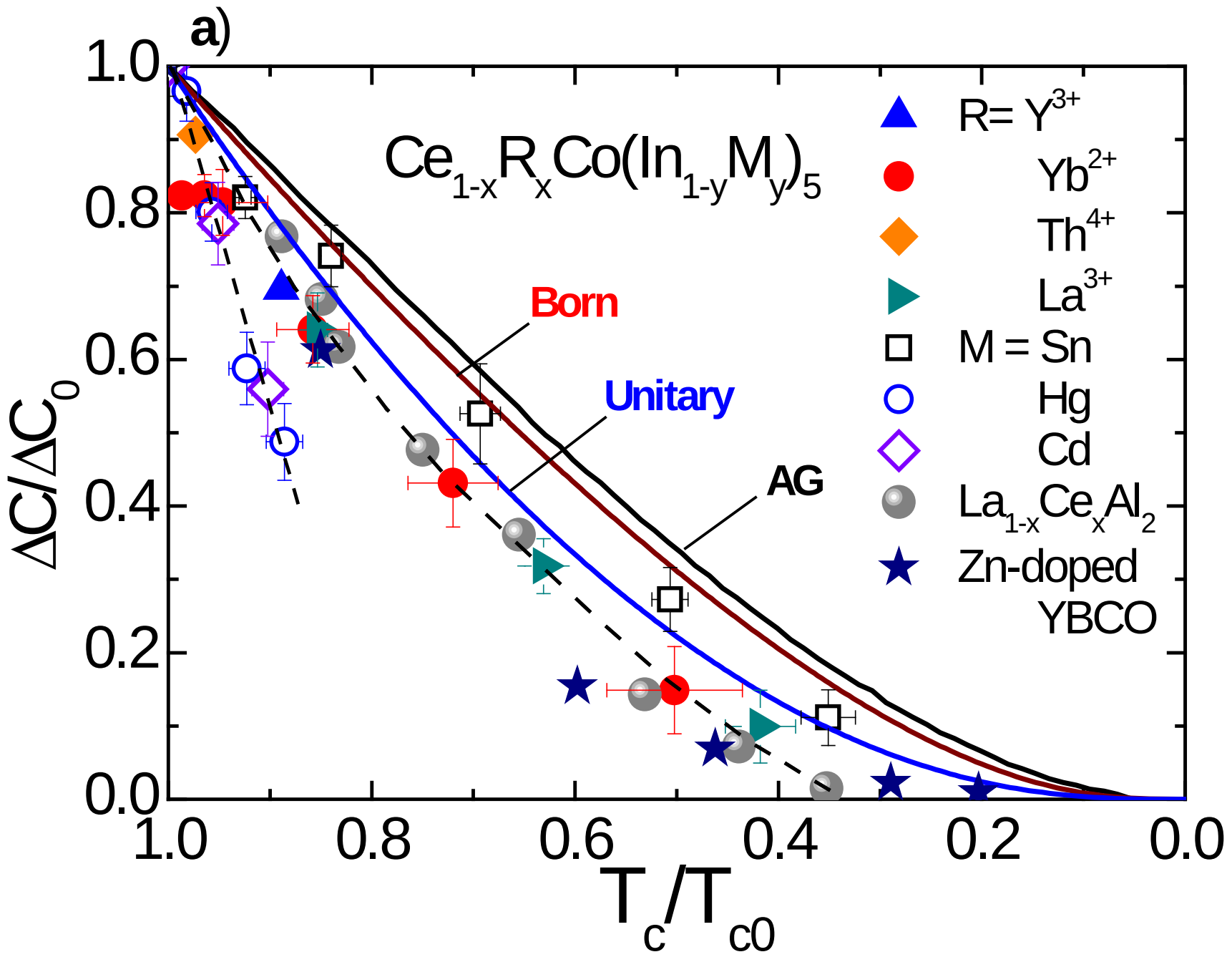

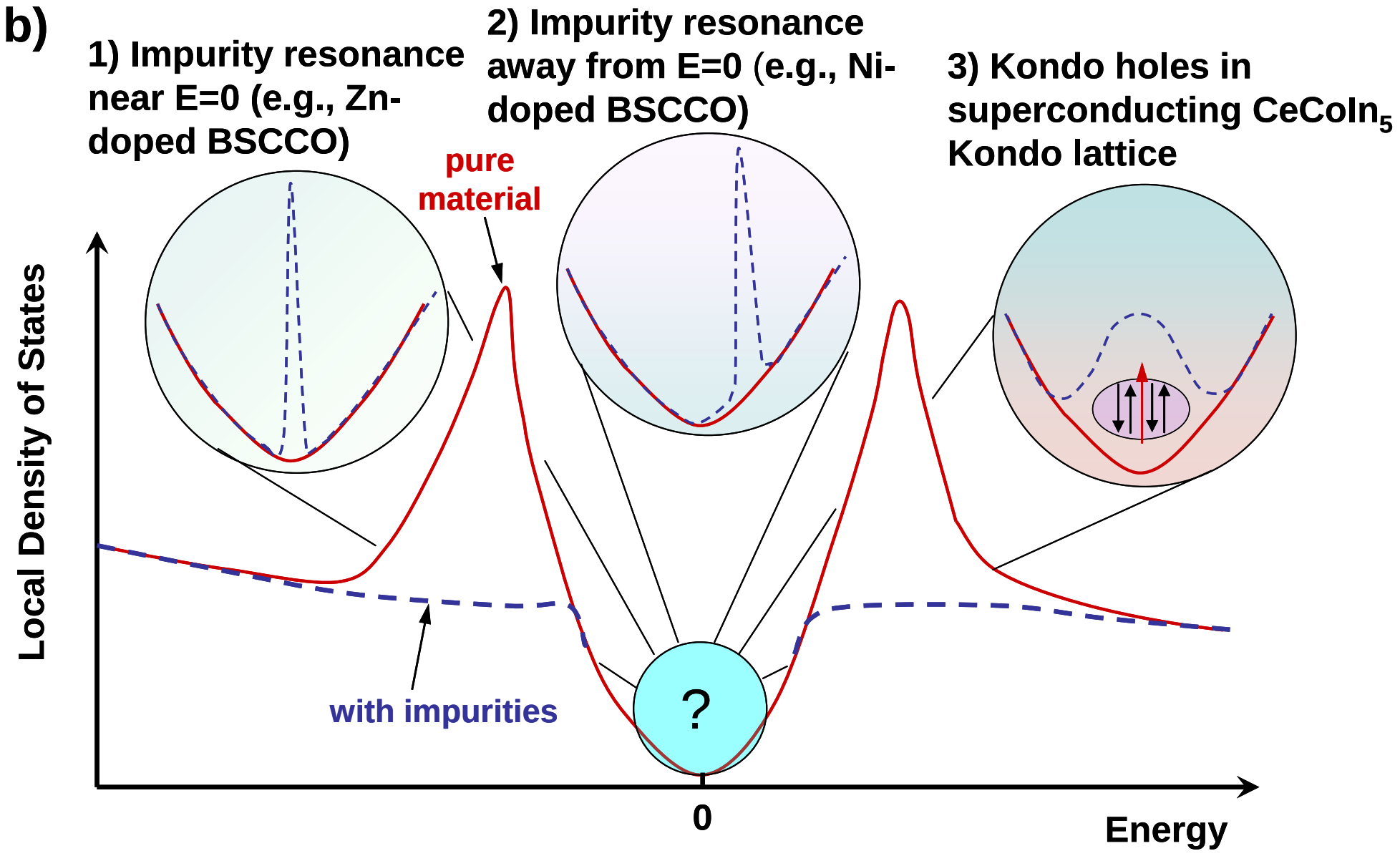
b)
1) Impurity resonance near E=0 (e.g., Zn-doped BSCCO)
2) Impurity resonance away from E=0 (e.g., Ni-doped BSCCO)
3) Kondo holes in superconducting $CeCoIn_5$ Kondo lattice
pure material
with impurities
?
Local Density of States
0
Energy

Fig. 3b

**Figure 4.** $^{115}$**In(1) Nuclear quadrupole resonance results in** $Ce_{1-x}La_xCoIn_5$ **for** $x$ **= 0, 0.1, and 1. a)** Spin-lattice relaxation rate, $1/T_1$, vs temperature T. The $1/T_1$ data for x = 1 is from Ref. (24) and for x = 0 from Ref. (25). **b)** $^{115}$In(1) NQR spectra for the highest transition (□7/2 □ □9/2) at T =3 K for x = 0 and 0.1 and at T = 0.4 K for x = 1. The solid lines are Gaussian (x = 0.1 and 1) and Lorentzian (x = 0) fits to the data. **c)** $^{115}$In(1) NQR spectra for x = 0.1 at several temperatures above and below $T_c$ ~ 0.9 K. The solid lines are Gaussian fits to the data. A similarly broadened peak was also observed in an $x$ = 0.05 sample but no other peaks were able to be resolved (not shown). The peaks A, B, C correspond to In(1) with 0, 1 and 2 nearest-neighbor (nn) La, respectively; the relative intensities of the main peak (0.7) and the two satellites are 0.2 and 0.09, reasonably close to the expected values for a simple binomial distribution (probabilities of 0.65 (0 nn), 0.29 (1 nn), and 0.05 (2 nn)), with a 10% chance of La occupying a Ce site. The error bars in **a)** were determined from Gaussian fits to the data in **c)**.

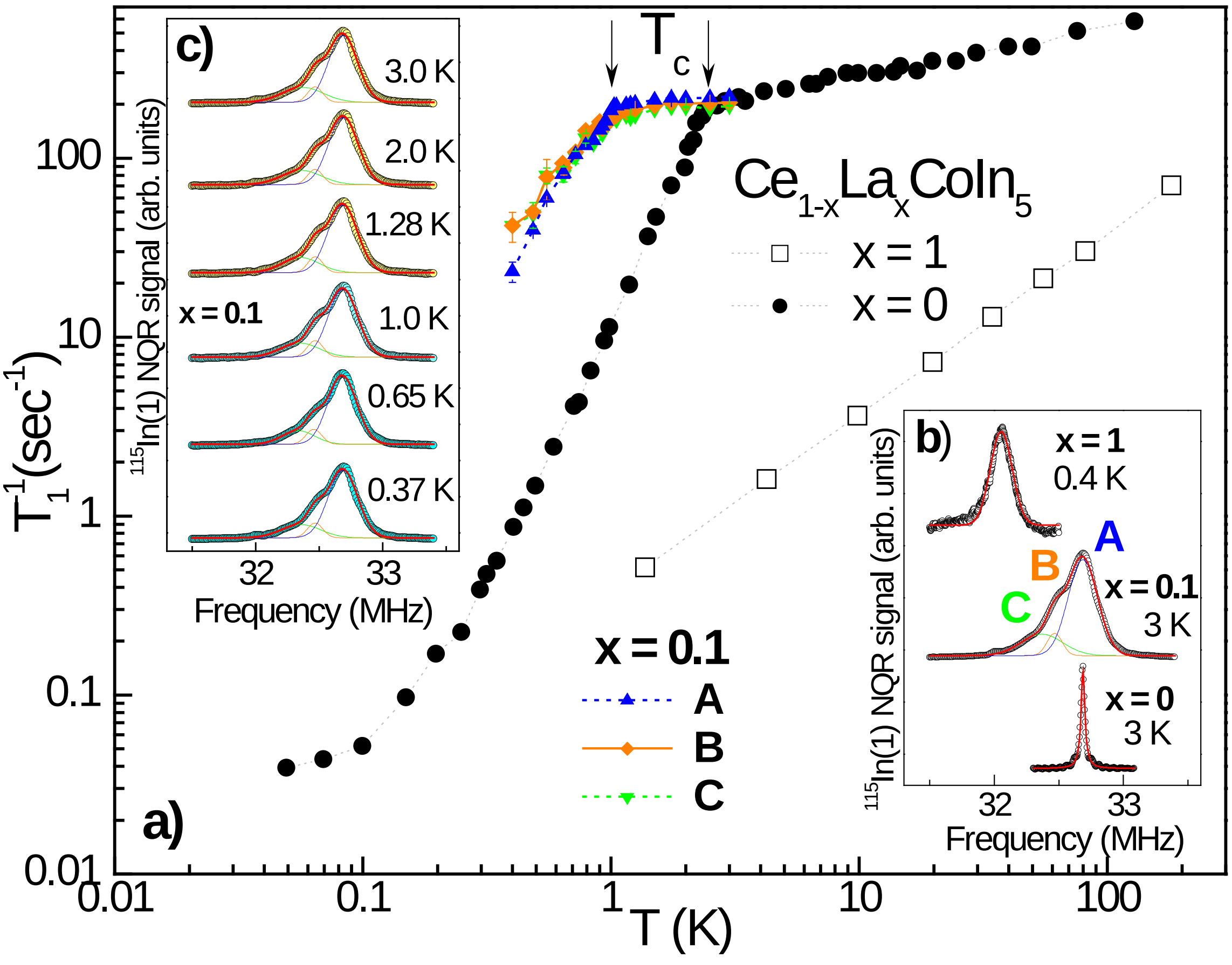

**Supporting Information**

Figure S1 shows the specific heat C/T and entropy S data of $Ce_{1-x}R_xCo(In_{1-x}M_x)_5$ (R = $Y^{3+}$, $Th^{4+}$; M = Hg). The three data sets were analyzed in a similar manner to data in Fig. 1, resulting in the normal state ($R_\gamma$) and superconducting state ($1-R_U$) fractions shown in Fig. 2b. The reduced specific heat jump $\Delta C/C_0$ vs $T_c/T_{c0}$ for these three impurities are displayed in Fig. 3a.

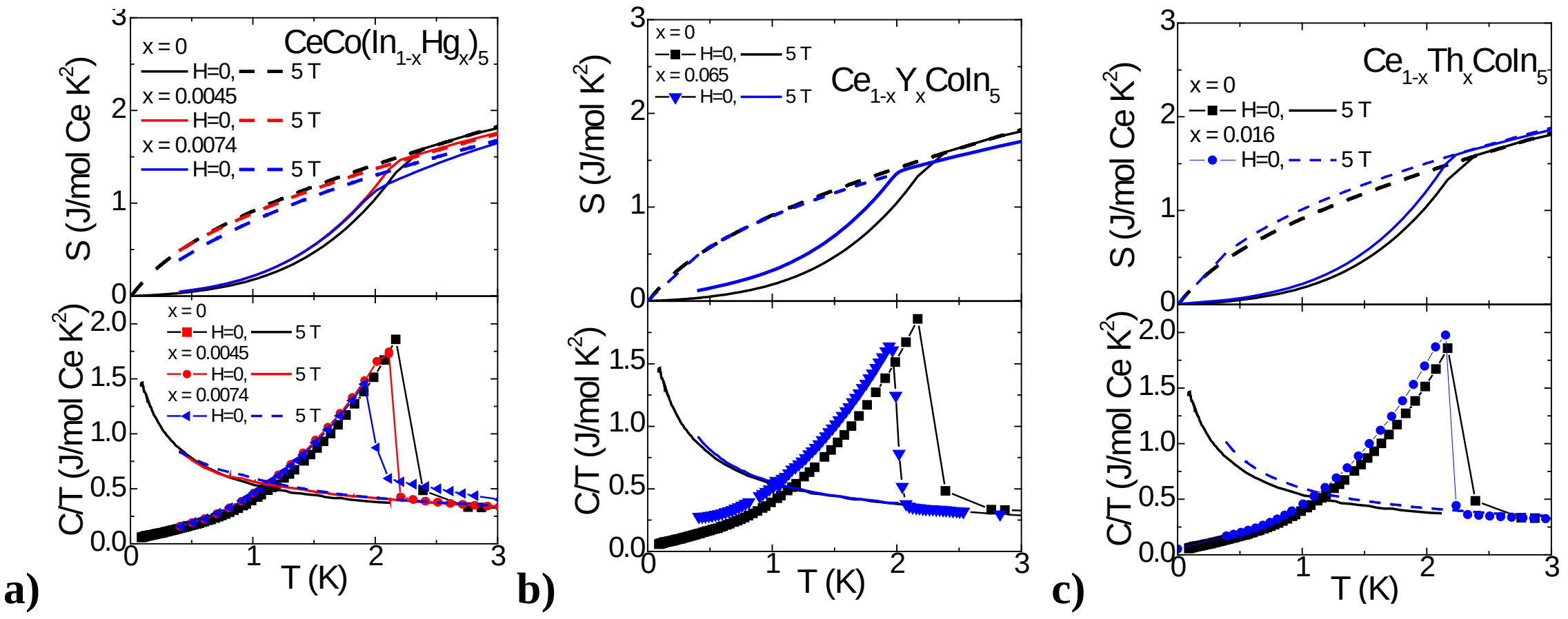

**Figure S1:** Specific heat divided by temperature C(T)/T and entropy S(T) in the superconducting and normal states of a) M = Hg, b) R=Y, and c) R=Th for $Ce_{1-x}R_xCo(In_{1-x}M_x)_5$.

Figure S2 shows the data of Fig. 2b along with calculations of the free energy of a d-wave superconductor in the presence of unitary impurity scattering, plotted as $\gamma_0/\gamma_N$ vs $1-U_{SC}(\Gamma/T_c)/U_{SC}(0)$, where $\Gamma$ is the impurity scattering rate, in agreement with Keller *et al.* [S1]. In addition to predicting a $x^{1/2}$ dependence of $\gamma_0$ as opposed to the linear dependence found experimentally (Fig. 2a inset), these calculations fail to reproduce the linear relation observed in the experimental data for all dopants in $CeCoIn_5$, $PuCoGa_5$, and $UPt_3$. These inconsistencies indicate that intra-gap states are not a result of conventional unitary or Born scatterers. Instead, these states suppress the superconducting order parameter in a manner similar to that of Zn-doped YBCO, providing further support for electronic inhomogeneity in these f-electron systems.

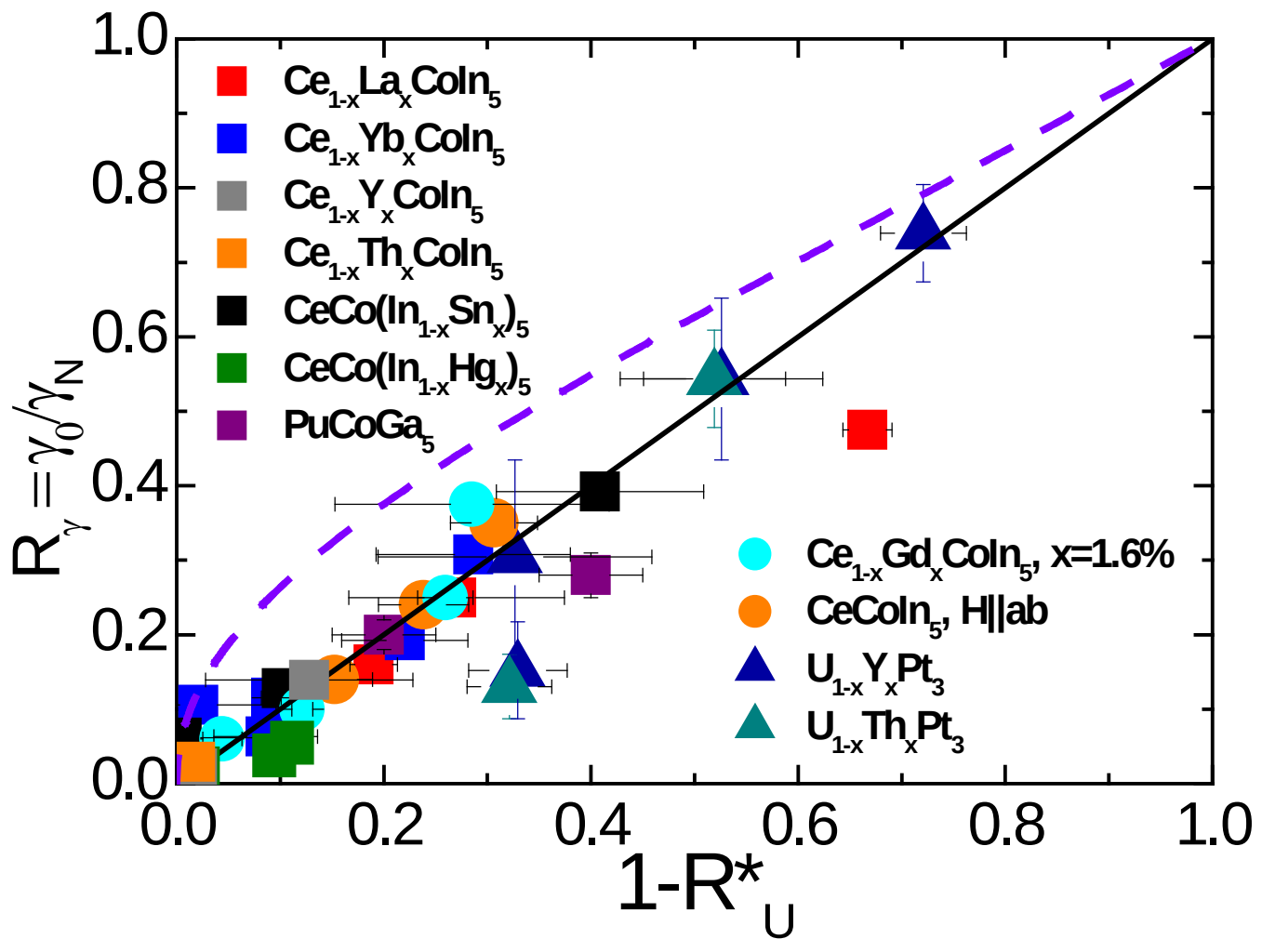

**Figure S2:** Normal state fraction of the inhomogeneous heavy fermion ground state $R_\gamma = \gamma_0/\gamma_N$ of $Ce_{1-x}R_xCo(In_{1-x}M_x)_5$ (R = $La^{3+}$, $Yb^{2+}$; M=Sn, Hg), $U_{1-x}M_xPt_3$ (M = Th, Y) and $PuCoGa_5$ (radiation damage induced impurities), $CeCoIn_5$ and $Ce_{0.994}Gd_{0.016}CoIn_5$ in magnetic field, and superconducting state fraction, $1 - R_U$, where $R_U = [U_{SC}(x)/T_c^2(x)]/[U_{SC}(0)/T_c^2(0)]$, obtained from the superconducting condensation energy $U_{SC} = \int_0^{T_c}(S_N - S_{SC})dT$. The dashed line is a free energy calculation for a d-wave superconductor with unitary impurity scattering, plotted as $\gamma_0/\gamma_N$ vs $1-U_{SC}(\Gamma/T_c)/U_{SC}(0)$, for arbitrary impurity scattering rate $\Gamma/T_c$, in agreement with Keller *et al.* [S1]. The solid line is a guide to the eye.

To obtain a reliable estimate of $\gamma_0$ in the superconducting state of $PuCoGa_5$, we use the measured 5f contribution to the specific heat for the “fresh” and aged samples combined with the reduction in the density of states determined from nuclear magnetic resonance data of similarly aged samples [S2]. Figure S3 displays the 5f contribution $\Delta C/T$ after the phonon contribution of $LuCoGa_5$ has been subtracted from the $PuCoGa_5$ data. A value of $\gamma_N = 68.5$ mJ/mol K$^2$, which conserves entropy between the normal and superconducting states, was used as an estimate since the normal state could not be reached in magnetic field (Fig. S3 inset). Figure S4a and S4b shows the temperature dependence of spin-lattice relaxation rate and Knight shift in the superconducting state of “fresh” and “aged” (~6 mo.) $PuCoGa_5$. The ratio of the density of states (DOS) of the superconducting to normal states is calculated from the following formulas:

$$\left(\frac{1}{T_1}\right) \Big/ \left(\frac{1}{T_1}\right)_{T=T_c} = \frac{2}{k_B T}\int \left\{ \left\langle \frac{N_s^2(E)}{N_0^2} \right\rangle + \left\langle M_s^2(E) \right\rangle \right\} f(E)[1 - f(E)]dE, \qquad (1)$$

$$K_s(T)/K_n = -\int_0^\infty \left\langle \frac{N_s(E)}{N_0} \right\rangle \frac{\partial f(E)}{\partial E} dE, \tag{2}$$

where $N_s(E) = \frac{N_0 E}{\sqrt{E^2 - \Delta^2(\theta,\ \phi)}}$, $M_s(E) = \frac{\Delta(\theta,\ \phi)}{\sqrt{E^2 - \Delta^2(\theta,\ \phi)}}$. Here, $N_0$ is the density of states, $f(E)$ is the Fermi distribution function, and $\langle \cdots \rangle$ represents an angular average over the Fermi surface. The superconducting gap was modeled as $\Delta(\theta,\ \phi,\ T) = \Delta_0(T) g(\theta,\ \phi)$ with a polar-type symmetry, $g(\theta,\ \phi) = \cos\theta$, where $\Delta_0(T)$ is assumed to have a BCS-type temperature dependence. As shown in Fig. S3a and S3b, the data are well-fit by Eqns. 1 and 2, respectively, yielding values of $N_{res}/N_0$ of 0.2 and 0.28 for "fresh" and "aged" samples, respectively. These values are not sensitive to the particular type of superconducting gap function (e.g., a two-dimensional type $g(\theta,\ \phi) = \cos(2\phi)$ ), which only results in a slightly modified zero-temperature gap value $\Delta_0(0)$. The residual Sommerfeld coefficient $\gamma_0$ is determined from these values of $N_{res}/N_0$ and the constant normal state value $\gamma_N$ above, for which the results are listed in Table 1. Using these values, the reduction in condensation energy (Fig. 2b) closely follows the behaviour to that of impurities introduced into $CeCoIn_5$ to create an inhomogeneous electronic state.

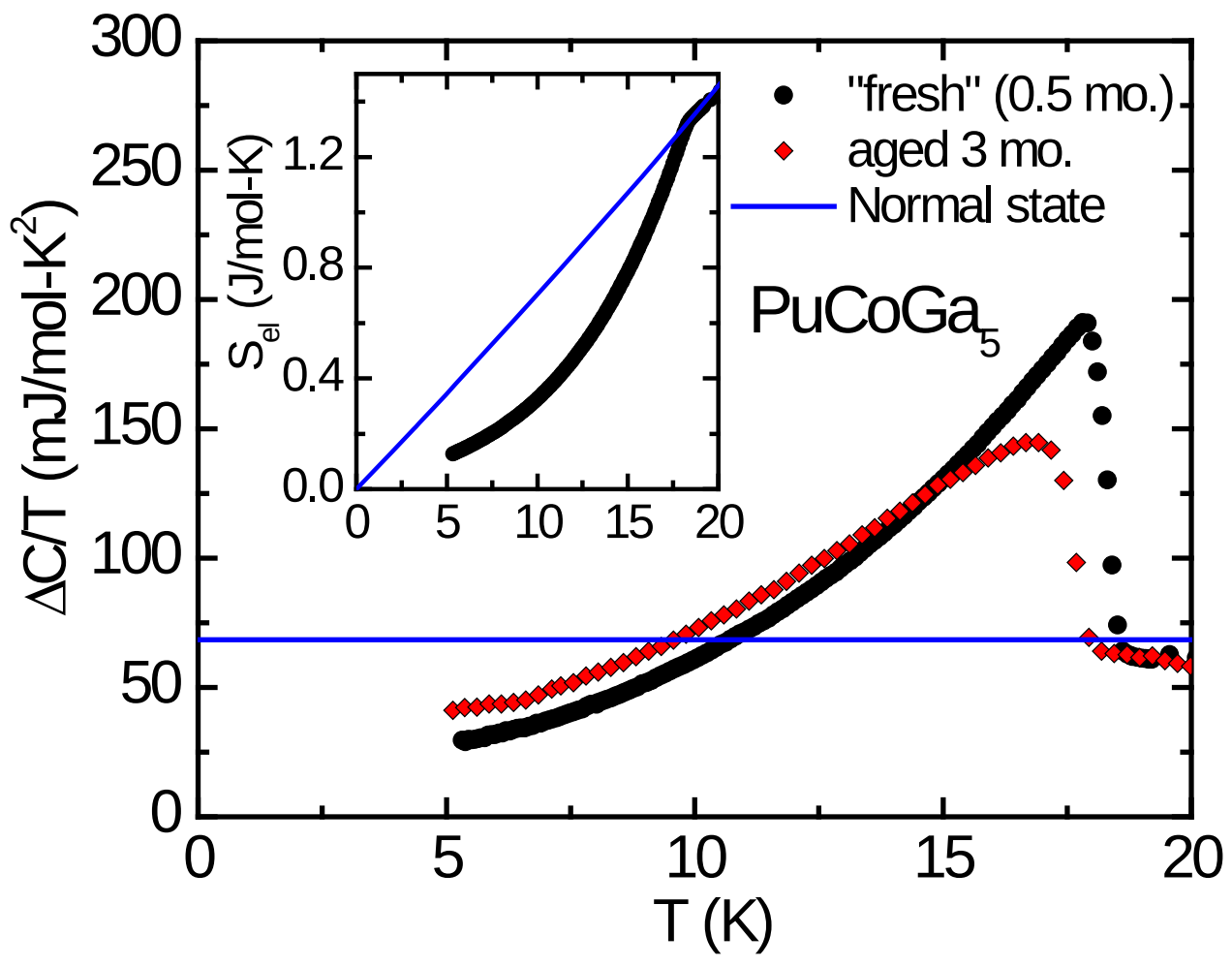

**Figure S3:** 5f contribution to the specific heat ΔC/T of $PuCoGa_5$ of "fresh" (2 weeks) and 3 month old sample, after accounting for the phonon contribution of nonmagnetic $LuCoGa_5$. The solid line is an estimate of the normal state contribution that conserves entropy between the normal and superconducting states. Inset: 5f contribution to the entropy $S_{el}$ of the "fresh" $PuCoGa_5$ sample and estimated normal state.

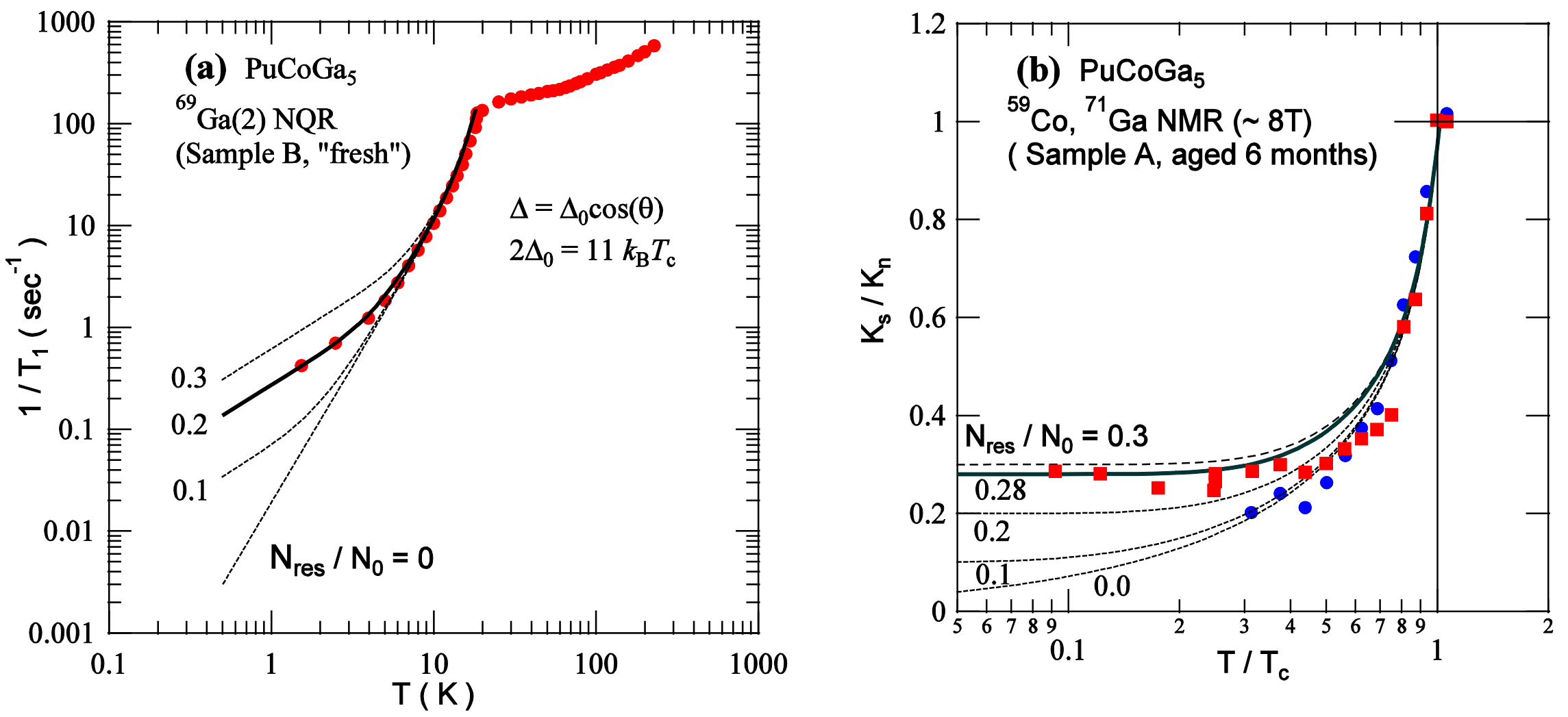

**Figure S4(a):** Spin-lattice relaxation rate of "fresh" $PuCoGa_5$ from Ref. [S2] showing the best fit to the data with a residual DOS $N_{res}$ to the normal state DOS $N_0$ of $N_{res}/N_0=0.2$. **(b)** Normalized Knight shift $K_s/K_n$ of $PuCoGa_5$ aged 6 months (roughly equivalent to 3 month old sample in Fig. S3) showing the best fit to the data with $N_{res}/N_0=0.28$.

**Table 1.** Parameters used in thermodynamic analysis of doped $Ce_{1-x}R_xCo(In_{1-x}M_x)_5$, $PuCoGa_5$, and $U_{1-x}M_xPt_3$ systems. x—impurity concentration, $T_c$, superconducting transition temperature determined from an equal area entropy conserving construction [K]; x, impurity concentration; $U_{sc}$ , superconducting condensation energy [J⁄mol-A, where A = Ce, Pu, U]; $\gamma_0$, zero-temperature electronic specific heat coefficient in superconducting state [J⁄mol-A $K^2$, where A = Ce, Pu, U]; $\gamma_N$, zero-temperature electronic specific heat coefficient in normal state [J⁄mol-A $K^2$, where A = Ce, Pu, U]; $1 - R_U$, normal fraction of superconducting matrix determined from the reduction in the superconducting condensation energy; $R_\gamma$, normal state fraction within superconducting state caused by impurities. The value of $E_c$ [in units of [J⁄mol-A $K^2$, where A = Ce, Pu, U] is also given in parentheses for each system.

| | $x$ | $T_C$ (K) | $U_{sc}$ (J/mol-Ce) | $U_{sc}/T_C^2$ (J/mol-Ce K$^2$) | $\gamma_0$ (J/mol-Ce K$^2$) | $\gamma_N$ (J/mol-Ce K$^2$) | $1-R_U$ | $R_\gamma$ |
|---|---|---|---|---|---|---|---|---|
| $Ce_{1-x}La_xCoIn_5$ | 0 | 2.25 | 1.18 | 0.238 | 0.04 | 2 | 0.020 | 0.02 |
| $E_c$=0.238 J/mol-Ce K$^2$ | 0.02 | 1.92 | 0.71 | 0.193 | 0.32 | 2 | 0.190 | 0.16 |
| | 0.05 | 1.42 | 0.35 | 0.174 | 0.5 | 2 | 0.270 | 0.25 |
| | 0.1 | 0.94 | 0.07 | 0.079 | 0.95 | 2 | 0.667 | 0.475 |
| | $x$ | $T_C$ (K) | $U_{sc}$ (J/mol-Ce) | $U_{sc}/T_C^2$ (J/mol-Ce K$^2$) | $\gamma_0$ (J/mol-Ce K$^2$) | $\gamma_N$ (J/mol-Ce K$^2$) | $1-R_U$ | $R_\gamma$ |
| $Ce_{1-x}Yb_xCoIn_5$ | 0.007 | 2.22 | 1.18 | 0.218 | 0.1 | 1.6 | 0.086 | 0.063 |
| $E_c$=0.238 J/mol-Ce K$^2$ | 0.02 | 2.17 | 1.076 | 0.233 | 0.19 | 1.8 | 0.021 | 0.106 |
| | 0.03 | 2.13 | 0.98 | 0.216 | 0.15 | 1.3 | 0.091 | 0.115 |
| | 0.073 | 1.93 | 0.691 | 0.185 | 0.25 | 1.3 | 0.220 | 0.192 |
| | 0.147 | 1.62 | 0.445 | 0.170 | 0.4 | 1.3 | 0.286 | 0.308 |
| $Ce_{1-x}Y_xCoIn_5$ | $x$ | $T_C$ (K) | $U_{sc}$ (J/mol-Ce) | $U_{sc}/T_C^2$ (J/mol-Ce K$^2$) | $\gamma_0$ (J/mol-Ce K$^2$) | $\gamma_N$ (J/mol-Ce K$^2$) | $1-R_U$ | $R_\gamma$ |
| $E_c$=0.238 J/mol-Ce K$^2$ | 0.065 | 2.00 | 0.78 | 0.208 | 0.26 | 1.9 | 0.128 | 0.139 |
| $CeCo(In_{1-x}Sn_x)_5$ | $x$ | $T_C$ (K) | $U_{sc}$ (J/mol-Ce) | $U_{sc}/T_C^2$ (J/mol-Ce K$^2$) | $\gamma_0$ (J/mol-Ce K$^2$) | $\gamma_N$ (J/mol-Ce K$^2$) | $1-R_U$ | $R_\gamma$ |
| $E_c$=0.238 J/mol-Ce K$^2$ | 0.006 | 1.85 | 0.81 | 0.237 | 0.12 | 1.95 | 0.003 | 0.062 |
| | 0.012 | 1.56 | 0.52 | 0.214 | 0.25 | 1.95 | 0.102 | 0.128 |
| | 0.024 | 0.8 | 0.09 | 0.141 | 0.51 | 1.3 | 0.409 | 0.392 |
| $CeCo(In_{1-x}Hg_x)_5$ | $x$ | $T_C$ (K) | $U_{sc}$ (J/mol-Ce) | $U_{sc}/T_C^2$ (J/mol-Ce K$^2$) | $\gamma_0$ (J/mol-Ce K$^2$) | $\gamma_N$ (J/mol-Ce K$^2$) | $1-R_U$ | $R_\gamma$ |
| $E_c$=0.238 J/mol-Ce K$^2$ | 0.0045 | 2.15 | 1.0 | 0.215 | 0.06 | 1.6 | 0.075 | 0.038 |
| | 0.0074 | 1.99 | 0.94 | 0.211 | 0.06 | 1.1 | 0.092 | 0.055 |
| $CeCoIn_5$ | *H* (*T*) | $T_C$ (K) | $U_{sc}$ (J/mol-Ce) | $U_{sc}/T_C^2$ (J/mol-Ce $K^2$) | $\gamma_0$ (J/mol-Ce $K^2$) | $\gamma_N$ (J/mol-Ce $K^2$) | $1\text{-}R_U$ | $R_\gamma$ |
| $E_c$=0.238 J/mol-Ce $K^2$ | 6 | 1.85 | 0.69 | 0.202 | 0.28 | 2 | 0.152 | 0.14 |
| | 8 | 1.39 | 0.35 | 0.179 | 0.48 | 2 | 0.238 | 0.24 |
| | 10 | 1 | 0.165 | 0.165 | 0.7 | 2 | 0.306 | 0.35 |
| | *H* (*T*) | $T_c$ (K) | $U_{sc}$ (J/mol-Ce) | $U_{sc}/T_C^2$ (J/mol-Ce $K^2$) | $\gamma_0$ (J/mol-Ce $K^2$) | $\gamma_N$ (J/mol-Ce $K^2$) | $1\text{-}R_U$ | $R_\gamma$ |
| $Ce_{1-x}Gd_xCoIn_5$ (x=0.016) | 0 | 2.28 | 1.27 | 0.252 | 0.12 | 2 | 0.044 | 0.06 |
| | 2 | 1.96 | 0.889 | 0.231 | 0.2 | 2 | 0.121 | 0.1 |
| $E_c$=0.26 J/mol-Ce $K^2$ | 3.5 | 1.52 | 0.452 | 0.196 | 0.5 | 2 | 0.259 | 0.25 |
| | 4.5 | 0.895 | 0.151 | 0.189 | 0.8 | 2 | 0.285 | 0.375 |
| $PuCoGa_5$ | *Age* (*Mo.*) | $T_C$ (K) | $U_{sc}$ (J/mol) | $U_{sc}/T_C^2$ (J/mol $K^2$) | $\gamma_0$ (J/mol $K^2$) | $\gamma_N$ (J/mol $K^2$) | $1\text{-}R_U$ | $R_\gamma$ |
| $E_c$=0.017 J/mol $K^2$ | 0.5 | 18.27 | 4.54 | 0.014 | 0.0137 | 0.069 | 0.2 | 0.2 |
| | 3 | 17.70 | 3.26 | 0.01 | 0.0192 | 0.069 | 0.4 | 0.28 |
| | *x* (*%*) | $T_C$ (K) | $U_{sc}$ (J/mol-U) | $U_{sc}/T_C^2$ (J/mol-U $K^2$) | $\gamma_0$ (J/mol $K^2$) | $\gamma_N$ (J/mol $K^2$) | $1\text{-}R_U$ | $R_\gamma$ |
| $U_{1-x}Y_xPt_3$ | 0 | 0.465 | 0.0145 | 0.0671 | 0.07 | 0.46 | 0.329 | 0.152 |
| $E_c$=0.1 J/mol-U $K^2$ | 0.08 | 0.395 | 0.0105 | 0.0674 | 0.14 | 0.46 | 0.326 | 0.304 |
| | 0.1 | 0.325 | 0.005 | 0.0474 | 0.25 | 0.46 | 0.526 | 0.543 |
| | 0.26 | 0.239 | 0.0016 | 0.0279 | 0.34 | 0.46 | 0.721 | 0.739 |
| $U_{1-x}Th_xPt_3$ | *x* | $T_C$ | $U_{sc}$ | $U_{sc}/T_C^2$ | $\gamma_0$ | $\gamma_N$ | $1\text{-}R_U$ | $R_\gamma$ |

| $E_c$=0.1 J/mol-U $K^2$ | (%) | (K) | (J/mol-U) | (J/mol-U $K^2$) | (J/mol $K^2$) | (J/mol $K^2$) | | |
|---|---|---|---|---|---|---|---|---|
| | 0 | 0.465 | 0.0145 | 0.0671 | 0.07 | 0.46 | 0.329 | 0.152 |
| | 0.17 | 0.39 | 0.0073 | 0.0481 | 0.25 | 0.46 | 0.519 | 0.543 |